\documentclass{moriond}

\usepackage{lineno}

\def\be{\begin{equation}}
\def\ee{\end{equation}}
\def\bea{\begin{eqnarray}}
\def\eea{\end{eqnarray}}

\newcommand{\pp}{p\kern-0.05em p}

\newcommand{\GeVc}{\ensuremath{\mathrm{GeV}\kern-0.05em/\kern-0.02em c}}

\newcommand{\pT}{\ensuremath{p_{\mathrm{T}}}}
\newcommand{\kT}{\ensuremath{k_{\perp}}}

\newcommand{\qhat}{\ensuremath{\hat{q}}}
\newcommand{\RAA}{\ensuremath{R_{\mathrm{AA}}}}

\begin{document}
\vspace*{4cm}
\title{Determining the jet transport coefficient $\hat{q}$ of the quark-gluon plasma using Bayesian parameter estimation}


\author{ James Mulligan$^{1,2}$ on behalf of the JETSCAPE Collaboration }
\address{$^1$Physics Department, University of California, Berkeley, CA 94720, USA \\ $^2$Nuclear Science Division, Lawrence Berkeley National Laboratory, Berkeley, California 94720, USA}

\maketitle

\abstracts{
We present a new determination of $\hat{q}$, the jet transport coefficient of the quark-gluon plasma.
Using the JETSCAPE framework, we use Bayesian parameter estimation to constrain the dependence of 
$\hat{q}$ on the jet energy, virtuality, and medium temperature from experimental measurements of inclusive hadron suppression in Au-Au collisions at RHIC and Pb-Pb collisions at the LHC. 
These results are based on a multi-stage theoretical approach to in-medium jet evolution with the MATTER and LBT jet quenching models. The functional dependence of $\hat{q}$ on jet energy, virtuality, and medium temperature is based on a perturbative picture of in-medium scattering, with components reflecting the different regimes of applicability of MATTER and LBT.
The correlation of experimental systematic uncertainties is accounted for in the parameter extraction. 
These results provide state-of-the-art constraints on $\hat{q}$
and lay the groundwork to extract additional properties of the quark-gluon plasma from jet measurements
in heavy-ion collisions.}

\section{Introduction}

At high temperatures, quantum chromodynamics (QCD) exhibits a deconfined state of matter
known as the quark-gluon plasma (QGP).\cite{Busza:2018rrf,Shuryak:2014zxa}
The nature of the degrees of freedom of the QGP, however, remains largely unknown. 
By studying this state, we seek to understand how complex behaviors arise from QCD,
including how strongly-coupled systems and their bulk properties emerge from quantum field theory.

Jets provide a compelling tool to pursue these questions. 
Depending on their transverse momentum (\pT{}) and substructure, 
jets can probe from the smallest medium scales to the largest medium scales, 
and jet evolution can be computed from first principles. 
Accordingly, a major experimental and theoretical jet physics program has developed
over the last two decades at the Relativistic Heavy Ion Collider (RHIC)
and the Large Hadron Collider (LHC), where ultrarelativistic heavy-ion
collisions create short-lived droplets of QGP.\cite{Qin:2015srf,Majumder:2010qh} 

This jet quenching program faces two major challenges. 
Firstly, jet evolution in heavy-ion collisions involves multiple stages of 
physics that are not known from first principles, such as the initial state of the collision, 
the hydrodynamic evolution of the medium, and the hadronization phase.
This can be addressed by global analyses that fit phenomenological models to a wide range 
of experimental data, and has recently been solved to a 
large extent.\cite{Everett:2020xug,Bernhard:2016tnd,Bernhard:2015hxa,Pratt:2015zsa,Nijs:2020roc} 
A second challenge, however, is that jet evolution in the QGP,
even for a perfectly characterized medium, involves
numerous theoretical unknowns, from 
the strength of coupling of the jet-medium interaction to the role of factorization breaking
and color coherence. No (known) golden observable exists to disentangle these open questions
about the jet-medium interaction. 
This necessitates global analyses of multiple jet observables.
Such global analyses provide a viable path not only to
disentangle various theoretical approaches describing the jet-medium interaction 
but also to precisely determine medium properties from experimental measurements --
the ultimate goal of studying the QGP.

In this work, we focus on the medium property known as the 
jet transverse momentum diffusion coefficient, \qhat,
which describes the average transverse momentum, $\langle \kT \rangle$, 
acquired by a parton as it traverses a given length $L$ of QGP. 
This transport coefficient is agnostic to the microscopic interactions that generate
the transverse diffusion, but rather characterizes the overall accumulated transverse momentum.
The transport coefficient \qhat{} has been calculated 
under certain approximations,\cite{Baier:1996sk,Zakharov:1997uu,Liou:2013qya,Blaizot:2014bha,Wu:2014nca,Arnold:2020uzm,Bianchi:2017wpt,Liu:2006ug} but in general involves nonperturbative
contributions that must be either computed on a lattice\cite{Majumder:2012sh,Kumar:2020wvb} or
extracted from experimental measurements.\cite{Burke:2013yra,Andres:2016iys,Chen:2010te,Ke:2020clc} 

We present a proof-of-principle determination of $\hat{q}$ 
using Bayesian parameter estimation -- the first such
extraction using Bayesian techniques -- laying the groundwork to extract additional 
properties of the quark-gluon plasma from jet measurements in heavy-ion collisions.\cite{Cao:2021keo}

\section{Analysis}

\begin{figure*}[!b]
\centering
\includegraphics[scale=0.75]{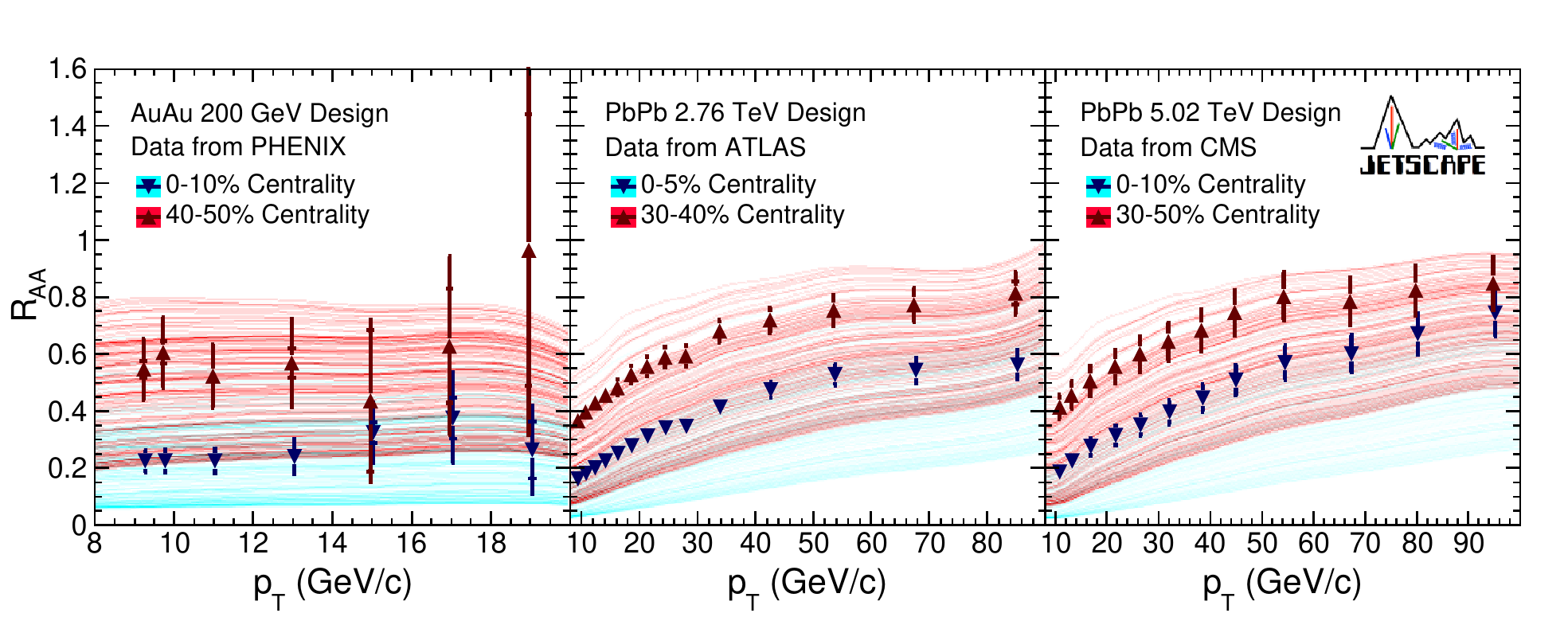} \\
\includegraphics[scale=0.75]{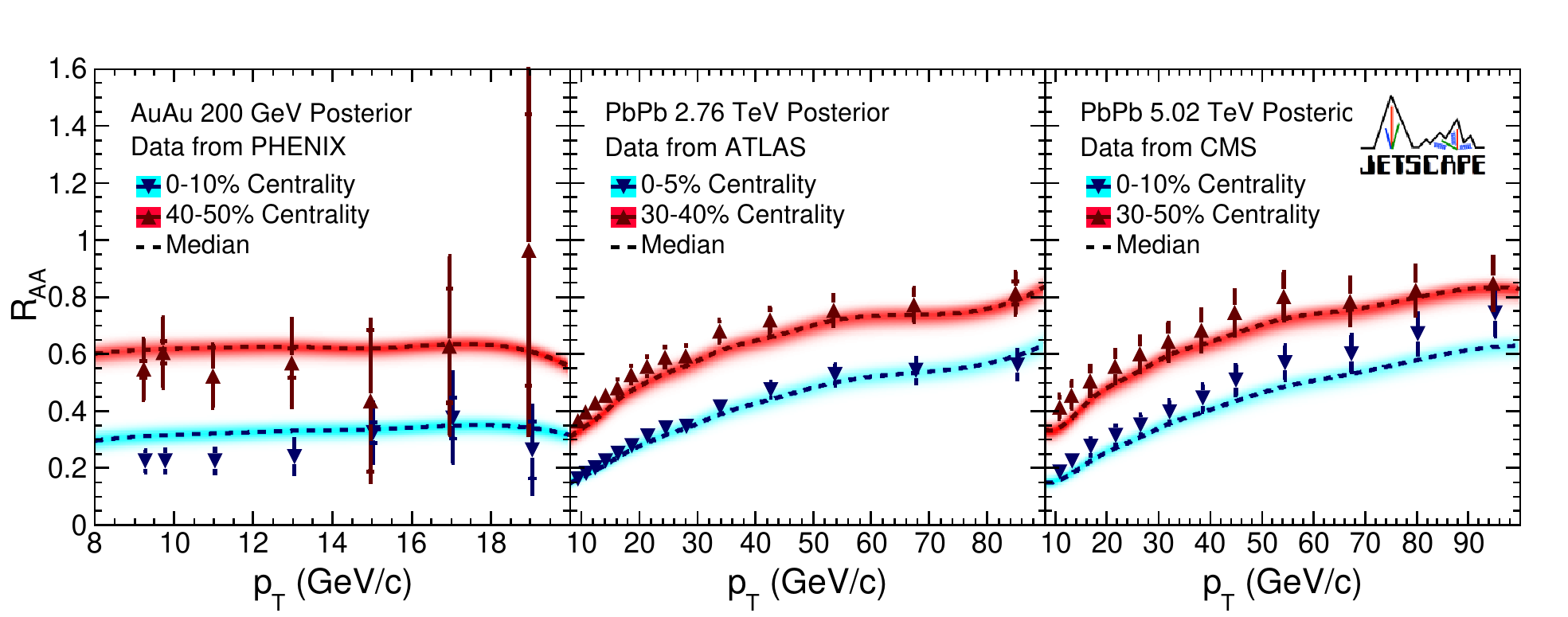}
\caption[]{Inclusive hadron \RAA{} for the three measured datasets,\cite{Adare:2012wg,CMS:2012aa,Aad:2015wga}
together with prior calculations (top) and posterior distributions (bottom) from the LBT model. 
Inner error bars on experimental data points are statistical errors; outer error bars are the quadrature sum 
of statistical error and systematic uncertainty.\cite{Cao:2021keo}}
\label{fig:posterior-raa}
\end{figure*}

We perform an extraction of \qhat{} using a selection of inclusive hadron \RAA{} data at 
RHIC and the LHC.\cite{Adare:2012wg,CMS:2012aa,Aad:2015wga}
We consider both central and semi-central data, shown in Fig. \ref{fig:posterior-raa}.
In order to properly assess the uncertainty correlations within and between the experimental measurements, 
we decompose the overall experimental covariance matrix into several sources, 
according to the varying degree of information reported by the experiments.\cite{Cao:2021keo}

We model the jet evolution using the JETSCAPE event generator framework\cite{Putschke:2019yrg,Cao:2017zih}
with the MATTER\cite{Majumder:2013re,Cao:2017qpx} and LBT\cite{Cao:2016gvr} jet quenching models.
MATTER is hypothesized to be valid in the high-virtuality, radiation-dominated regime,
whereas LBT is hypothesized to be valid in the low-virtuality, scattering-dominated regime.
We additionally consider a multi-stage model in which high-virtuality partons evolve according to MATTER,
and low-virtuality partons evolve according to LBT, which we denote ``MATTER+LBT''.
Within these models, we parameterize \qhat{} as a function of the medium temperature, $T$, 
and parton energy, $E$, with an ansatz consisting of a sum of a high-virtuality, 
$T$-independent term, and a low-virtuality, elastic scattering term:
\be 
\label{eq:1}
\frac{\qhat\left(E,T\right) |_{\theta=\{A,B,C,D\}}}{T^3}=42C_R\frac{\zeta(3)}{\pi}\left(\frac{4\pi}{9}\right)^2\left\{\frac{A\left[\ln\left(\frac{E}{\Lambda}\right)-\ln(B)\right]}{\left[\ln\left(\frac{E}{\Lambda}\right)\right]^2}+\frac{C\left[\ln\left(\frac{E}{T}\right)-\ln(D)\right]}{\left[\ln\left(\frac{ET}{\Lambda^2}\right)\right]^2}\right\},    
\ee
where $\theta\equiv\{A,B,C,D\}$ are parameters that will be determined from the experimental data using Bayesian parameter estimation.\cite{Kennedy_2001}
For the multi-stage MATTER+LBT model, we consider two adaptations of this parameterization
in which an additional parameter characterizing the virtuality switching scale between
the two models is included.\cite{Cao:2021keo}

Starting with broad prior distributions of $\theta$, we use Bayesian parameter estimation
to produce posterior probability distributions of $\theta$, 
and thereby $\qhat\left(E,T\right)$.
To do so, we employ Gaussian Process Emulators to interpolate our computationally expensive
event generator across $\theta$-space, and at each explored $\theta$ evaluate the likelihood
to observe the experimental data given the model results at that particular $\theta$. 
The posterior distributions are sampled using Markov Chain Monte Carlo. 
By using Bayesian parameter estimation, this procedure improves upon previous extractions of \qhat{}
due to its rigorous statistical approach and quantification of uncertainties on
the extracted \qhat{}.
Further details can be found in a recently submitted article.\cite{Cao:2021keo}

\begin{figure*}[!b]
\centering
\includegraphics[scale=0.5]{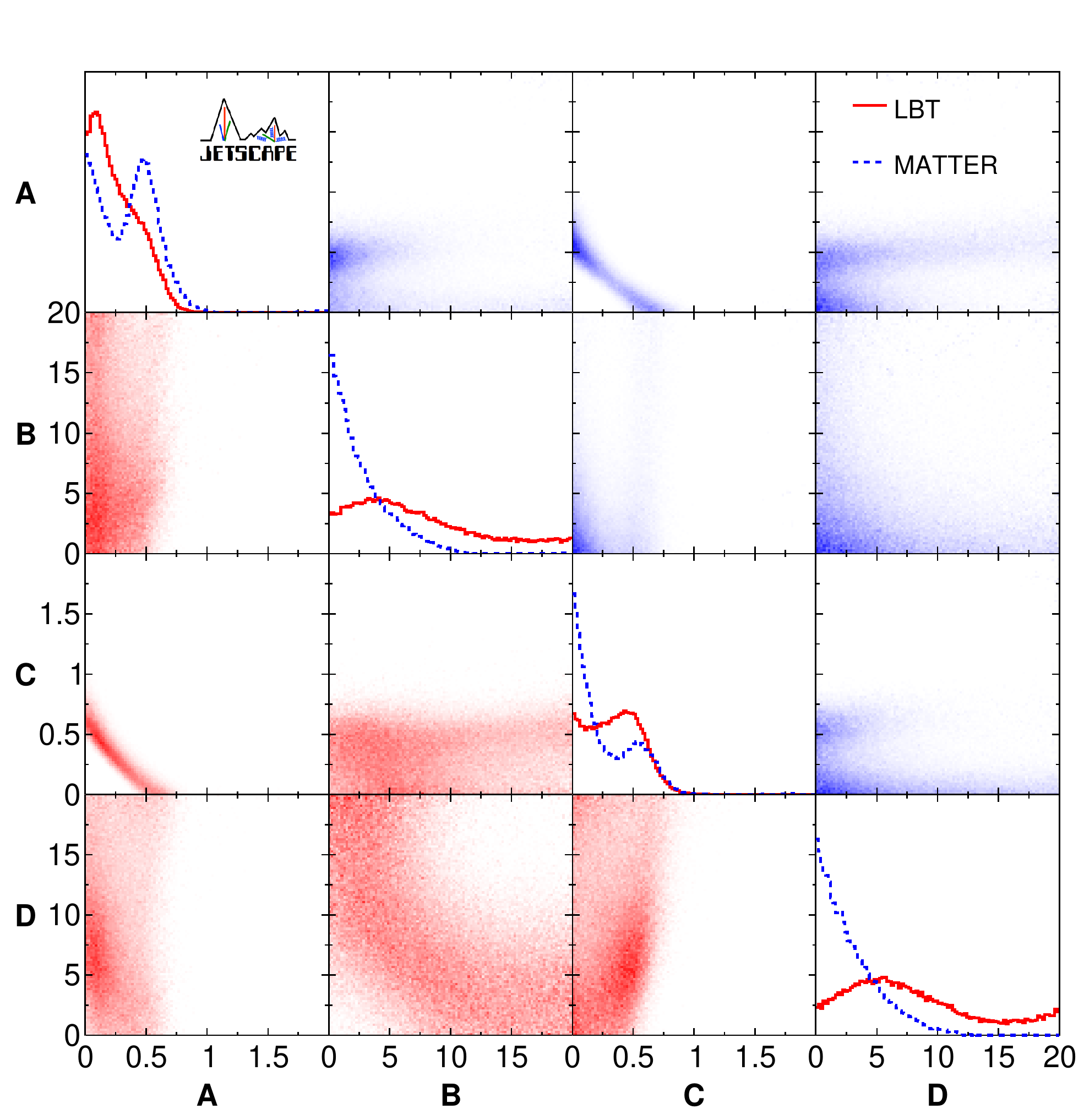}
\caption[]{Posterior distributions of the $\theta\equiv\{A,B,C,D\}$ space for \qhat{} 
when MATTER and LBT are applied separately. 
Off-diagonal panels show correlations of posterior distributions for LBT (lower left, red) 
and MATTER (upper right, blue).\cite{Cao:2021keo}}
\label{fig:posterior-theta}
\end{figure*}

\section{Results}

We perform separate extractions of \qhat{} using either MATTER, LBT, or MATTER+LBT. 
Figure \ref{fig:posterior-raa} shows an example of the explored \RAA{} values before (top) 
and after (bottom) constraining the model to the experimental data.
We find that LBT describes the data reasonably well, with some small systematic deviations,
as does MATTER, albeit with slightly larger deviations. 
For MATTER+LBT, we find no evidence that the multi-stage model improves the description of the
experimental data, suggesting that further theoretical work is needed.

The posterior distributions of $\theta\equiv\{A,B,C,D\}$ are shown in Fig. \ref{fig:posterior-theta}
for both LBT and MATTER. 
We find that the extracted parameters are substantially different for LBT compared to MATTER.
In particular, LBT exhibits a preference for smaller values of $A$ and larger values of $C$, 
whereas MATTER exhibits a preference for larger values of $A$ and smaller values of $C$.
This is in fact consistent with the original motivation of the \qhat{} parameterization
ansatz in Eq. \ref{eq:1}: The first additive term, associated with high-virtuality physics, has 
overall coefficient $A$ -- and is preferred by the radiation-dominated MATTER model --
while the second additive term, associated with elastic scattering off of a thermal medium, 
has overall coefficient $C$ -- and is preferred by the scattering-dominated LBT model.

\begin{figure*}[!b]
\centering
\includegraphics[scale=0.39]{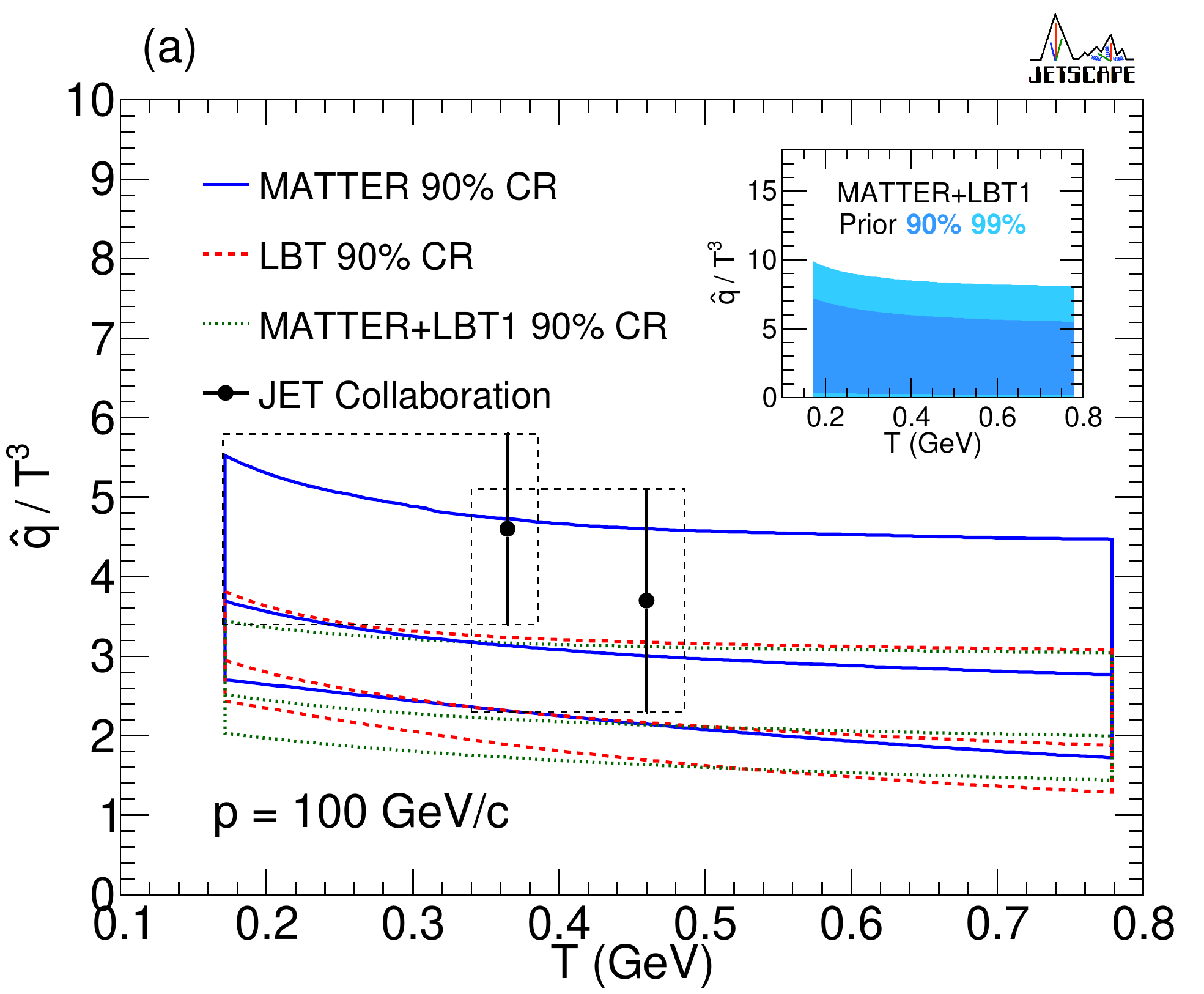}
\includegraphics[scale=0.39]{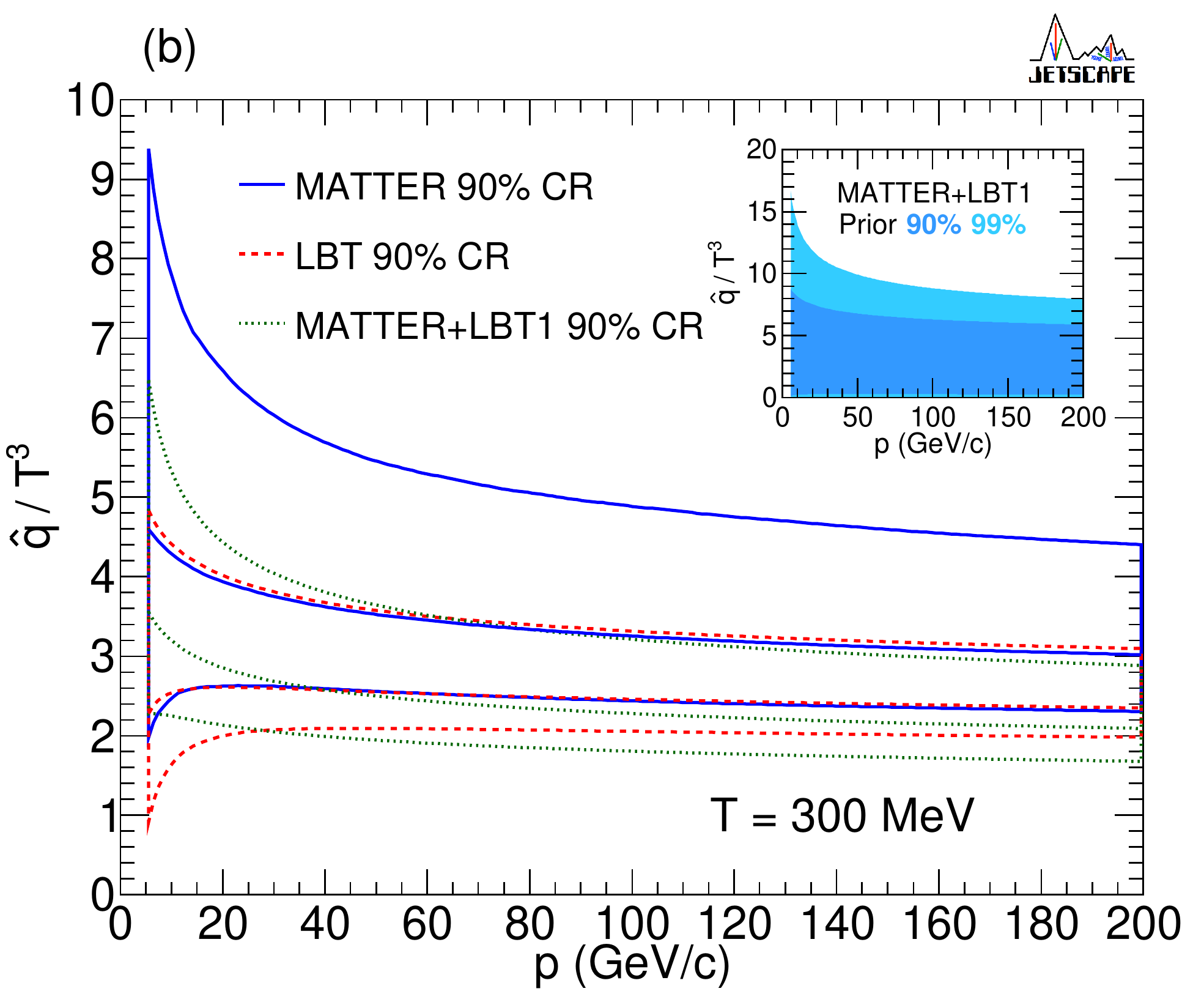}
\caption[]{The (quark) jet transport coefficient \qhat{} from Bayesian parameter extraction using 
MATTER and LBT separately as well as MATTER+LBT: (a) as function of the medium temperature, and (b) as function of quark momentum. 
The lines at the center of the bands indicate their median values.\cite{Cao:2021keo}}
\label{fig:posterior-qhat}
\end{figure*}

From these parameter posterior distributions, we plot the extracted $\qhat/T^3$ 
as a function of the medium temperature $T$ and parton momentum $p$ in Fig. \ref{fig:posterior-qhat}.
We plot the the prior distributions of $\qhat/T^3$ in the insets of Fig. \ref{fig:posterior-qhat},
which demonstrate that the data provide considerable constraints on the value of \qhat{}.
We find a generally weak dependence on both $T$ and $p$, 
consistent with earlier work by the JET Collaboration.\cite{Burke:2013yra}
The values of \qhat{} are similar between the different models considered, although
with notably smaller central values in the multi-stage MATTER+LBT model, 
due to the fact that quenching is performed over a wider range of parton virtualities
in the multi-stage model than in MATTER or LBT alone.

\begin{figure*}[!t]
\centering
\includegraphics[scale=0.46]{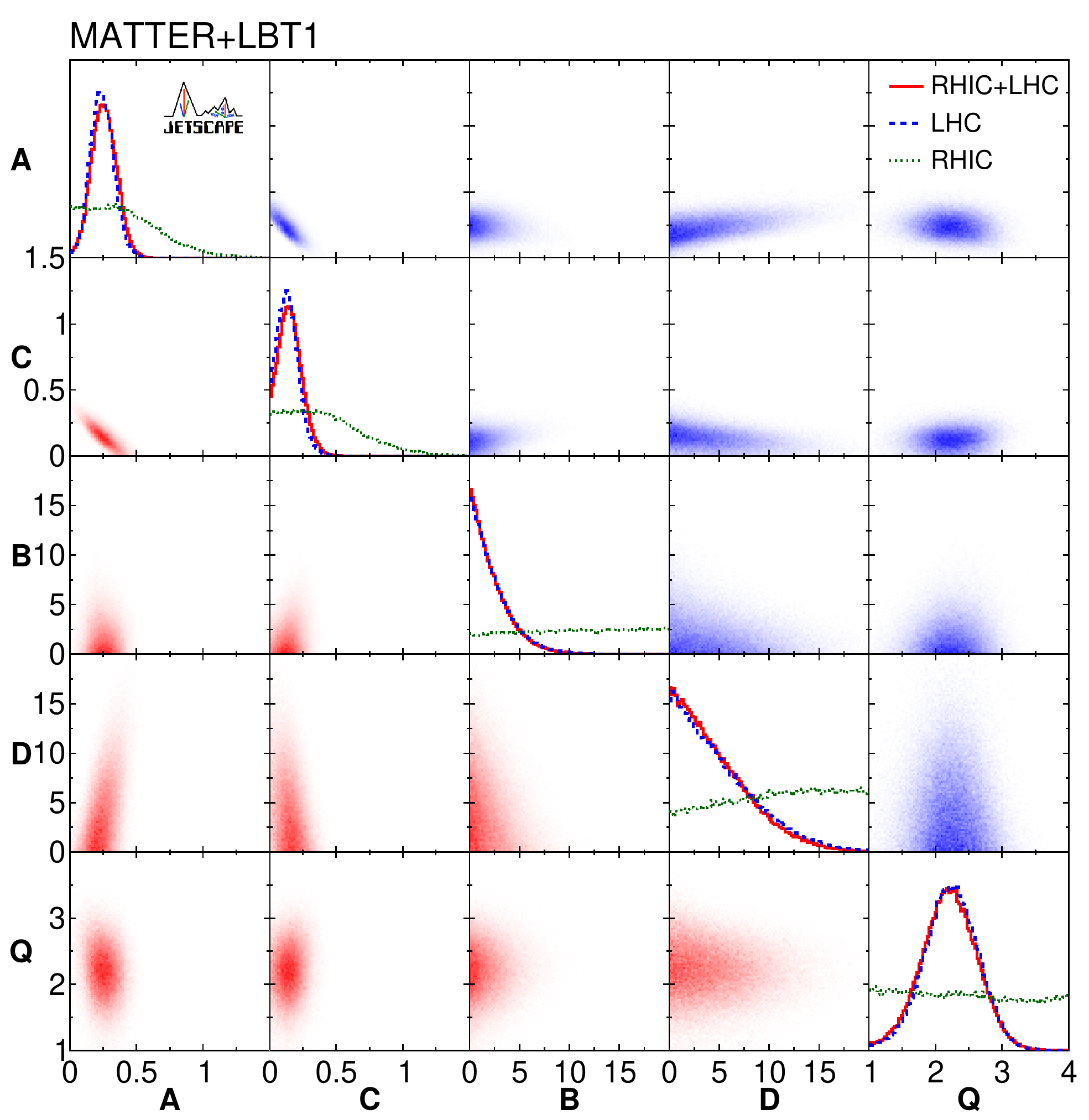}
\caption[]{Posterior distribution
of $\theta\equiv\{A,B,C,D\}$ with the multi-stage MATTER+LBT model, 
for separate extractions using either RHIC, LHC, or RHIC+LHC data. 
Off-diagonal panels show correlations of posterior distributions for 
RHIC+LHC (lower left, red) and LHC only (upper right, blue).\cite{Cao:2021keo}}
\label{fig:posterior-rhic-lhc}
\end{figure*}

Bayesian parameter estimation can also be used to study the impact of particular observables
on the precision of the extracted quantities. 
Figure \ref{fig:posterior-rhic-lhc} shows the impact of RHIC vs. LHC data by separately
extracting \qhat{} with each collider dataset.
We find that the posterior distributions are dominated by the LHC data.
This can be partially attributed to the impressive precision and scope of the LHC measurements, 
but we also note that it is impacted by our choice of input data, which for this analysis
we limited to $\pT > 8 \;\GeVc$, thus intrinsically disfavoring RHIC data. 
This calls for future study, and highlights the important role that Bayesian parameter
estimation can play in guiding experimental measurements. 

\section{Summary}

We extracted the jet transverse diffusion coefficient, \qhat{}, of the quark-gluon plasma
as a continuous function of the medium temperature and parton momentum.
By using Bayesian parameter estimation, we improved upon previous extractions of \qhat{}
with rigorous statistical methods and quantification of uncertainties.
We explored several models within the JETSCAPE event generator framework: 
MATTER, LBT, and a multi-stage model MATTER+LBT.
We used a set of hadron \RAA{} measurements at RHIC and the LHC, and found
that the data provides significant constraints on the prior distributions,
with results consistent with previous extractions.

Global analysis will be key to uncovering the nature of deconfined QCD matter.
This study serves as a proof-of-principle that can be systematically 
extended to include additional observables,
such as fully reconstructed jets, and extraction of 
additional medium properties, such as the path-length dependence of jet quenching,
and eventually the nature of the constituents of deconfined QCD.

\section*{Acknowledgments}

This work was supported in part by the National Science Foundation (NSF) within 
the framework of the JETSCAPE collaboration, under grant number ACI-1550228.

\section*{References}
{
\footnotesize
\typeout{}
\bibliography{references}
}

\end{document}